# Experimental investigation to influence of pump phase fluctuation on phase-correlation of output optical fields from a non-degenerate parametric oscillator

**Dong Wang, Yana Shang, Zhihui Yan, Wenzhe Wang, Xiaojun Jia, Changde Xie[*], and Kunchi Peng**

*State Key Laboratory of Quantum Optics and Quantum Optics Devices,*
*Institute of Opto-Electronics of Shanxi University, Taiyuan 030006, P.R.China*

[*]*Corresponding author: changde@sxu.edu.cn*

The influence of the phase fluctuation of the pump laser on the phase-correlation between the signal and idler modes of the output fields from as non-degenerate optical parametric oscillator operating above oscillation threshold was experimentally investigated. The noise spectra of the intensity-difference and the phase-sum of the entangled optical beams were measured with a pair of unbalanced fiber Match-Zehnder interferometers specifically designed. The experimental results proved the previously theoretical prediction and are in reasonable agreement with the calculation based on semiclassical theory involving the phase fluctuation of pump laser.

Optical parametric oscillators both degenerate (DOPO) and nondegenerate (NOPO) have been one of the most important entanglement sources in the quantum information of continuous variables (CV) [1]. As a useful tool in quantum optics for the generation of nonclassical states of light, the quantum fluctuation characteristics of the output fields from OPO has been extensively studied since the late 1980s. Firstly M. D. Reid and P. D. Drummond theoretically predicted that the signal and idler optical modes produced from NOPO operating above or below the oscillation threshold are in an entanglement state with the quantum correlation of amplitude and phase quadratures [2-5]. In 1992, the CV entanglement of optical modes was experimentally demonstrated by Z. Y. Ou et al. with an NOPO below threshold [6]. Although the quantum correlation of amplitude quadratures between intense signal and idler from above-threshold NOPO (named twin beams) was measured experimentally and effectively applied by several groups [7-12], the phase correlation of the twin beams with nondegenerate frequencies was not observed for a long time. For measuring the phase correlation of twin beams using general homodyning detector, Laurat et al. forced the NOPO to oscillate in a strict frequency-degenerate situation by inserting a $\lambda/4$ plate inside the NOPO and observed a 3 dB phase-sum variance above the shot-noise-limit (SNL) [13]. Then, in 2005 CV entanglement of frequency-nondegenerate twin beams was experimentally measured by two groups with scanning a pair of tunable ring analysis cavities [14] or with two sets of unbalanced Match-Zehnder (M-Z) interferometers [15], respectively. In 2006 the entangled twin beams with stable frequency difference was obtained by Pfister's group [16]. However, those experimentally measured phase-correlations were always smaller than that of the theoretical prediction. At the end of 1980s, Fabre et al. presented the expressions for calculating quantum correlations of amplitude and phase quadratures between twin beams [17], in which the pump laser was regarded as an ideal coherent state thus the influence of the pump excess phase noise on entanglement of twin beams was not considered. In the Fig. 3 of Ref. [15] achieved by our group before, we compared the normalized noise power spectra of the intensity-difference and the phase-sum calculated from the expressions given in Ref. [17] and the experimentally measured values, which showed that the measured value of the intensity-difference perfectly agreed with that calculated theoretically but the measured phase correlation noise was higher by 0.08 than the calculated values. Earlier, Reid and Drummond theoretically proved that a correlation between signal and idler phase diffusion existed and hence that the sum-phase fluctuations are not affected by this noise [4]. To explain why the experimentally measured phase-correlation of twin beams were always less than the theoretical values, the effect of the excess phase noise of the pump laser was considered in the theoretical calculations recently [18-21]. All calculations show that the excess phase noise of the pump laser will reduce the phase-correlation of twin beams and no more effect to the intensity-correlation. So far, to the best of our knowledge there is no experimental demonstration on this effect. In this paper we present the first experimental study on the effect of the pump excess phase noise to the phase-correlation of twin beams. For directly comparing with theory, we derived a formula of the intensity-difference ($S_p$) and the phase-sum ($S_q$) noise spectra between output signal and idler of NOPO with the semiclassical method used in Ref. [17], but in our calculation the excess phase noise of the pump laser, $E(\omega)$, is involved:

$$S_p(\omega) = 1 - \eta \frac{TT'}{T'^2 + \omega^2 \tau^2} \tag{1}$$

$$S_q(\omega) = 1 - \eta' \frac{TT'}{T'^2 \sigma^2 + \omega^2 \tau^2} + \eta' \frac{2TT'(\sigma - 1)}{T'^2 \sigma^2 + \omega^2 \tau^2} E \tag{2}$$

where $\eta$ and $\eta'$ are the transmission efficiency for the intensity and phase correlation measurement, respectively; $T'=T+\delta$, $T$ is the transmission coefficient of the output mirror of the NOPO and $\delta$ is the extra intracavity loss in the NOPO; $\sigma=(P/P_0)^{1/2}$ is the pump parameter ($P$—the pump power, $P_0$ — the threshold pump power of the NOPO); $\omega$ is the analysis circular frequency in the range of radio frequency (rf); $\tau$ is the round trip flying time of photons in the optical cavity. The detailed theoretical analyses on this subject will be presented in another publication which is being prepared. The Eqs. (1) and (2) were deduced under the condition of that the finesse of the NOPO cavity for the pump laser is much smaller than the finesse for the signal and idler beams as that used in Ref. [17].

The measurement scheme and principle of the phase fluctuation using an unbalanced M-Z interferometer are the same as that presented by Glöckl et al. originally [22]. According to the requirement in Ref. [22], for measuring the phase fluctuation of the input field of the interferometer the difference in length between the short and the long arms of the unbalanced interferometer ($\triangle L$) should equal to $c/(2\omega n)$, where ω is the analysis circular frequency, c is the light velocity in vacuum and n is the refraction index of transmission medium. As that shown in Eqs. (1) and (2), the quantum fluctuation of the intensity-difference and the phase-sum are decreased when the analysis circular frequency (ω) is lower for a given NOPO. It means that the quantum correlation of twin beams is higher at lower analysis frequency. For measuring the phase correlation of twin beams at the different analysis frequency conveniently we constructed a pair of unbalanced fiber M-Z interferometers with totally equivalent elements. The length of the short arm of the interferometers is fixed at 2 m and that of the long arm can be selected among 50 m, 21 m and 12 m, which corresponds to the requirement for the analysis frequencies of 2 MHz, 5 MHz and 10 MHz, respectively. The refraction index of the fiber equals to 1.55. In order to study quantitatively the influence of the pump excess phase noise on the phase-correlation of twin beams, a phase modulator is placed in the optical path of the pump laser before NOPO. The given modulation intensity stands for a certain excess white phase noise of pump laser.

Fig. 1 is the schematic of the unbalanced M-Z fiber interferometer which consists of a polarizing-beam-splitter (PBS), two reflection mirrors ($M_1$ and $M_2$), and a 50/50 optical beam-splitter ($M_3$). The short and long fibers with input and output fiber couplers (FC) constitute the short and long arms of the interferometer. The half wave plates $HWP_1$ and $HWP_2$ are used for aligning the polarization direction of light relative to PBS and that between two arms respectively. The output optical fields from $M_3$ are detected by a balanced detection system consisting of high-efficiency photodiodes ($D_1$ and $D_2$). The sum and difference of the photocurrents detected by $D_1$ and $D_2$ can be achieved with the positive and negative power combiners (+/-), respectively. If aligning the polarization of input light with HWP1 to make total input light only passes through the short fiber, in this case, $M_3$, $D_1$ and $D_2$ play a role of general balanced detector, thus the sum and difference photocurrents give the amplitude noise level and the SNL of the input light, respectively [15, 22]. If splitting the input light equivalently to the two arms of the interferometer and adjusting the phase difference between the two optical beams passing through the short and long arms to $\pi/2$, the difference photocurrent corresponds to the phase noise level of the input light and the sum photocurrent stands for the SNL [15, 22]. The measured losses of both short and long arms are about 22% which mainly come from the losses of the fiber couplers. The loss in the fiber can be neglected, thus the losses in two arms are almost equivalent.

The experimental system is shown in Fig. 2. The pump laser at 540 nm wavelength is produced from a homemade CV frequency-doubled and frequency-stabilized Nd:YAP/KTP laser (Nd:YAP—Nd-dopped YAlO$_3$ perovskite, KTP—polassium titanyl phosphate)[23]. The semi-monolithic NOPO consists of an α-cut type-II phase-matching KTP crystal and an output coupler which is a concave mirror of 50 mm curvature with high reflectivity at 540 nm and the transmission coefficient of 3.2% at 1080 nm. The front face of KTP is coated to be used as the input coupler of the NOPO with the reflectivity of 39% at 540 nm and high reflectivity at 1080 nm. The other face is coated with the antireflective film at both wavelengths at 540nm and 1080nm to reduce the intracavity loss. The output coupler is mounted on a piezoelectric transducer (PZT) for actively locking the cavity length of the NOPO on resonance with the pump laser. The measured cavity finesse for 1080nm is 150, and the total intracavity extra loss is about 1%. The twin beams with cross-polarized directions are produced in the NOPO through a nonlinearly optical process of frequency down-conversion. The output signal and idler beams are separated by the polarizing-beam-splitter (PBS), and then each of them is directed into an unbalanced M-Z interferometer (MZI$_1$ and MZI$_2$). The difference and the sum photocurrents are analyzed and recorded by the spectrum analyzer (SA). In the system, three spectrum analyzers are used totally, two of them in MZI$_1$ and MZI$_2$ and another one for the final measurements of the photocurrent combinations [15].

During the experiment for all measurements the pump power of the NOPO is kept at 194 mW, which is 64 mW higher than the oscillation threshold power of ～130 mW. Although the phase correlation should be better for the pump power close to the threshold according to the theoretical calculation [2，7], we found in experiments that the NOPO with lower finesse of the pump laser could operate stably only when a higher pump power was used. The highly unstable output from the NOPO was observed when the pump power approached the threshold. We believe that the reason for these results is that the multiple modes can oscillate simultaneously in the NOPO with low finesse of the pump laser owing to its flat resonance peak, thus mode competition and mode hoping must induce instability. When the pump power is increased, once the oscillation of a twin beam mode dominates in the NOPO, the output will be stable. Under the pump power of 194mW the detected output power of the twin beams is about 20 mW.

In the experiment, at first the amplitude and phase noise spectra of the signal and idler modes are measured by $MZI_1$ and $MZI_2$, respectively. Then, the quantum correlations of twin beams are defined by the noise level of the intensity difference and the phase sum of the photocurrents measured by the two interferometers. Fig. 3(a), (b) and (c) are the measured quantum correlation variances of the signal and idler beams at 2 MHz, 5 MHz and 10 MHz, respectively, in which three different lengths (50 m, 21 m and 12 m) of the long arms of two interferometers are used respectively to meet the requirement of the phase-correlation measurement at three different analysis frequencies as mentioned before. In the three figures, trace i, v and iv correspond to the SNL of the twin beams, the noise powers of the intensity-difference and the phase-sum under the situation without adding excess phase noise on the pump laser, respectively. When excess white phase noise of 0.33 or 1.00 (normalized to the original phase noise of the pump laser) is added on the pump laser through a phase-modulator, the phase-correlation noise power increases from iv to iii (0.33) or ii (1.00). As that predicted by theory [17-21], the intensity-correlation v is not effected by the excess phase noise of the pump laser. All amplitude and phase correlation variances measured at the three analysis frequencies are lower than the SNL even with the excess pump phase noise of 1.00, thus the quantum entanglement of twin beams experimentally proved. However, we found in experiment that when the excess pump phase noise was larger than 3.0, the phase-correlation noise power will be higher than the SNL. The critical excess pump phase noise can be calculated from Eq. (2). The results perhaps can explain why the phase-correlation of twin beams in some experiments was not observed [13].

Fig. 4 shows the normalized noise power spectra of the intensity-difference and the phase-sum of twin beams calculated from Eqs. (1) and (2) by use of the parameters of the real experimental systems( τ =39.5ns, σ =1.39, $\eta$=0.9*0.78 and $\eta'$=0.9*0.78*0.78 coming from the fiber coupling efficiency of 78% and the detection efficiency of 90%). Since the input optical beam of MZI passes through two fibers (short and long arms) in the measurement of the phase noise, we have $\eta'$=0.9*0.78*0.78, but it only transmits through the short arm for the measurement of the intensity-difference, thus $\eta'=0.9*0.78$. Trace i corresponds to the intensity-difference spectrum. Trace ii, iii, iv corresponds to the phase-sum spectra with the pump excess noise E = 0, 0.33, 1.00 respectively. The symbols ▲, ▼, ► and ◄ are the experimental measured values corresponding to the intensity-difference and the phase-sum with E=0, 0.33 and 1.00, respectively. The experimental analysis frequencies are 2MHz, 5MHz and 10MHz. We can see that the experimental measurements are in reasonable agreement with that theoretically calculated. Still, the experimentally measured values of the phase-sum noise (▲, ►, ◄) are a little higher than that of theory (ii, iii, iv). We consider that is because the possibly spurious phase-noise of the pump laser produced inside the NOPO has not been considered in Eq. (2), which will be studied in our next work.

In conclusion, we experimentally demonstrated the dependence of the phase-correlation of twin beams from NOPO operating above threshold on the excess phase noise of the pump laser. The experimental measurements agree with the theoretical predictions resulting from the semiclassical method quite well. The intense frequency-nondegenerate entangled light beams probably are appropriate to be used for quantum key distribution of optical continuous variables since the local oscillations for homodyne detection will not be needed and also for distributing quantum information between different parts of the electromagnetic spectrum by means of teleportation.

This research was supported by the PCSIRT (Grant No. IRT0516), Natural Science Foundation of China (Grants No. 60608012，No. 60608012, 60736040 and 10674088).

## References without titles

## List of figure captions

Fig. 1 Schematic of the fiber M-Z interferometer.

PBS: Polarization beam splitter; FC: fiber coupler; $HWP_{1-2}$: half wave-plate; telescope: lenses combination; $M_{1-2}$: mirror with 100% reflectivity; M3: 50/50 beam splitter; $D_{1-2}$ : photo-detector; +/-: positive and negative power combiners; SA: spectrum analyzer

Fig. 2 Experimental set-up.

PM: phase modulator; $MZI_{1-2}$: M-Z interferometer; +/-: positive and negative power combiner; SA: spectrum analyzer.

Fig.3 Noise spectrum of quantum correlations at 2 MHz (a), 5 MHz (b) and 10 MHz (c).

i: SNL; ii: the phase-correlation noise power with 1.00 white noise modulation; iii: the phase-correlation noise power with 0.33 white noise modulation; iv: the noise power of the phase-correlation; v: the noise power of the intensity-difference.

Fig.4 Normalized correlation noise. Trace i (▼) corresponds to the theoretical (experimental) values of the amplitude-difference. Trace ii (▲), trace iii (►) and trace iv (◄) correspond to the theoretical (experimental) values of the phase- sum with E = 0, 0.33, 1.00, respectively.

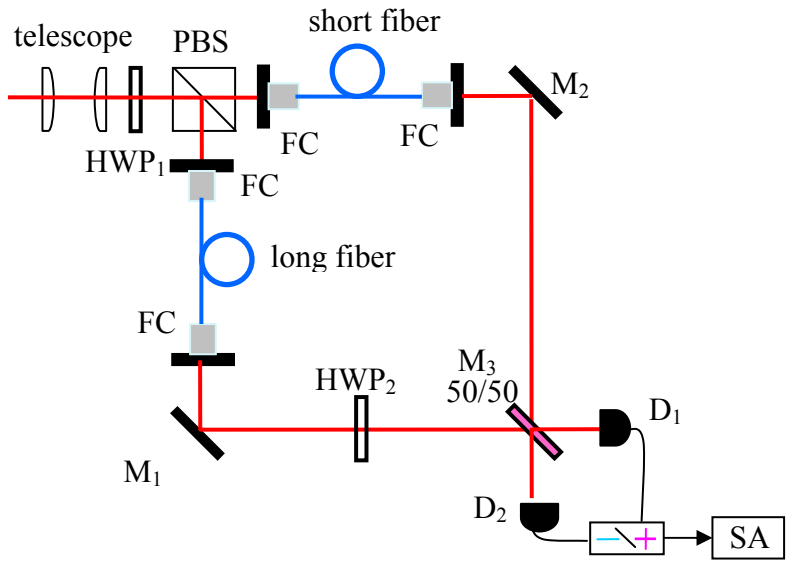

Fig. 1

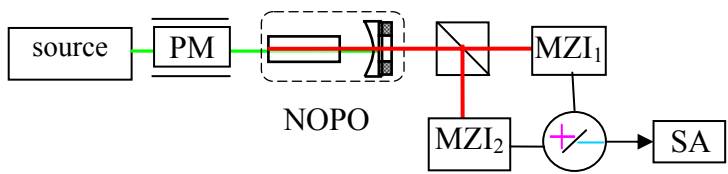

Fig. 2

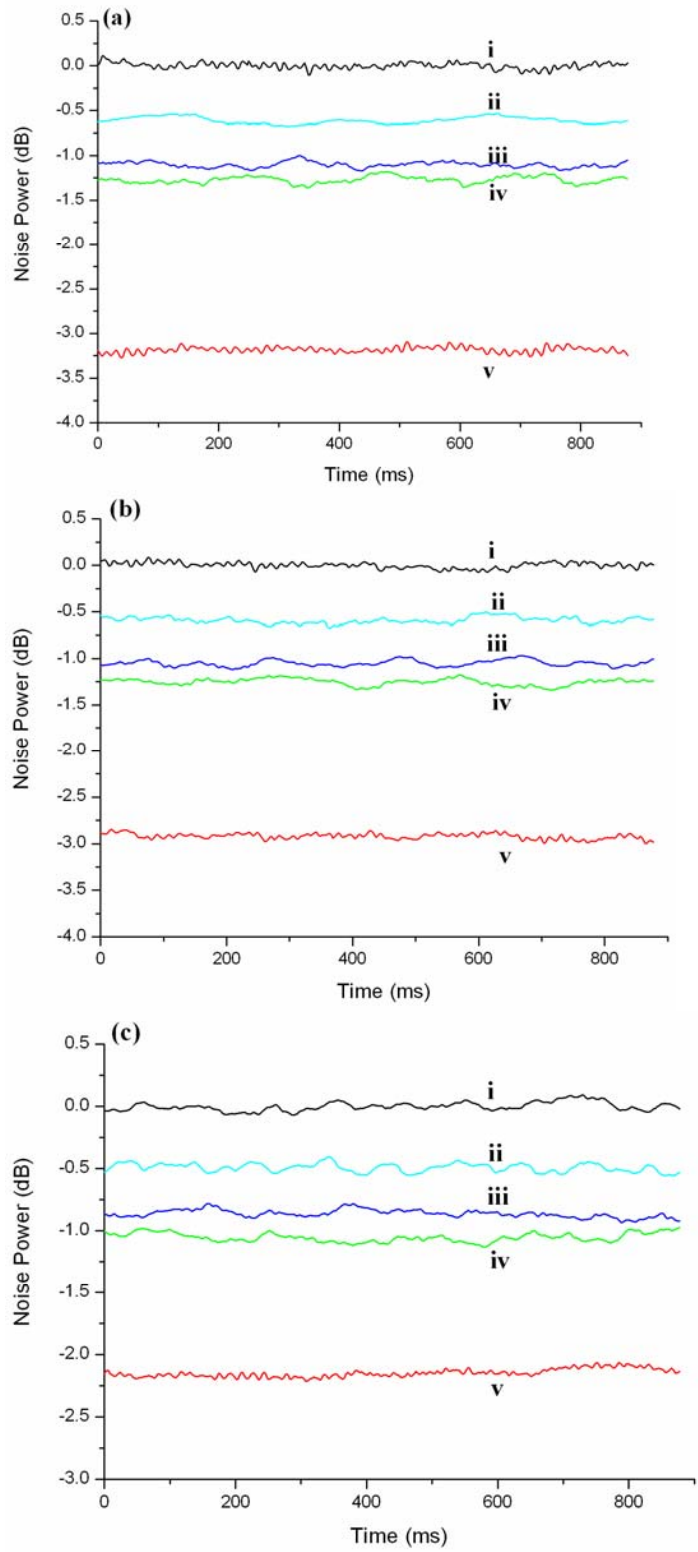

Fig. 3

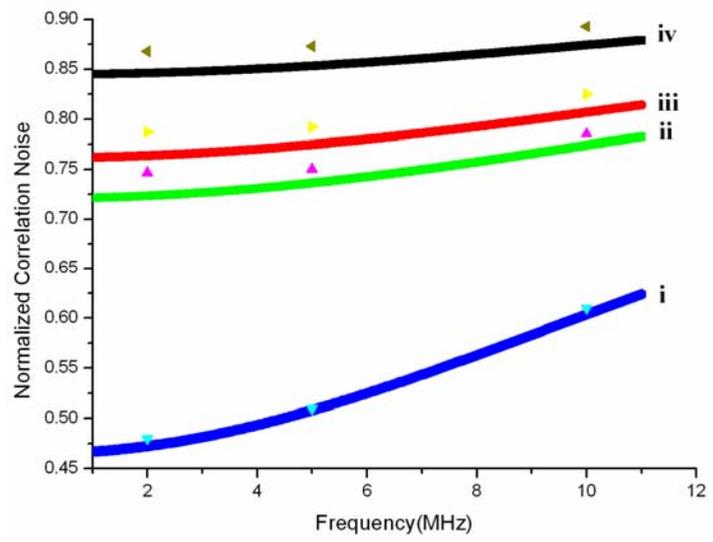

Fig. 4